 \documentclass[preprint,
                preprintnumbers,
                amsmath,
                amssymb,
                superscriptaddress]{revtex4}

\usepackage{latexsym}
\usepackage{amsfonts}
\usepackage{amsmath}
\usepackage[dvips]{graphicx}
\usepackage{color}
\usepackage{bm}

\newcommand{\placefig}[1]{
    \begin{center}
     {\bf FIGURE \ref{#1} AROUND HERE}
    \end{center}
}

\newcommand{\placetab}[1]{
    \begin{center}
     {\bf TABLE \ref{#1} AROUND HERE}
    \end{center}
}

\def\figfoot{Vendrell et. al., Journal of Chemical Physics}

\newcommand{\figcaption}[2]{
    \noindent {\bf Figure \ref{#1}:} #2
    \vspace{1cm}
}

\def\icm{cm$^{-1}$}
\def\wat{\mbox{H$_2$O}}
\def\zun{\mbox{H$_5$O$_2^+$}}

\def\hyd{\mbox{H$_3$O$^+$}}
\def\asymH{\mbox{[O-H-O$_\parallel$]}}
\def\symOO{\mbox{[O-O$_\parallel$]}}

\def\ESwagA{106}      
\def\ESwagAaa{108}      
\def\ESwagAab{254}      
\def\ESwagB{232}      
\def\ESwagC{374}      
\def\ESwagD{422}      

\def\ESrotO{1}        
\def\ESrotA{103}      
\def\ESrotB{126}      
\def\ESrotC{210}      

\def\ESrock{481}      
\def\ESrockB{943}     %
\def\ESrockC{930}     %
\def\ESrockD{915}     %

\def\ESRA{550}        
\def\ESRB{1069}       

\def\ESdoubl{918}     
\def\ESdoubh{1033}    
\def\ESzB{2338}        

\def\ESperpP{1391}     

\def\ESzARA{1411}     
\def\ESzARB{1898}     

\def\ESbendIP{1606}   
\def\ESbendOP{1741}   

\def\ESstrSIP{3607}   
\def\ESstrSOP{3614}   
\def\ESstrA{3689}     

\begin{document}

\author{Oriol Vendrell}
\email[e-mail: ]{oriol.vendrell@pci.uni-heidelberg.de}
\affiliation{Theoretische Chemie, Physikalisch-Chemisches Institut, 
   Universit\"at Heidelberg, INF 229,
   D-69120 Heidelberg, Germany}

\author{Fabien Gatti}
\email[e-mail: ]{gatti@univ-montp2.fr}
\affiliation{%
  LDSMS (UMR 536-CNRS), CC 014, Universit{\'e} de Montpellier II,
  F-34095 Montpellier, Cedex 05, France
}

\author{Hans-Dieter Meyer}
\email[e-mail: ]{Hans-Dieter.Meyer@pci.uni-heidelberg.de}
\affiliation{Theoretische Chemie, Physikalisch-Chemisches Institut, 
   Universit\"at Heidelberg, INF 229,
   D-69120 Heidelberg, Germany}

\title{Full Dimensional (15D) Quantum-Dynamical Simulation of the Protonated Water-Dimer II:
       Infrared Spectrum and Vibrational Dynamics}

\date{\today}

\begin{abstract}
   The infrared absorption spectrum of the protonated water dimer
   (H$_5$O$_2^+$) is simulated in full dimensionality (15D) in
   the spectral range 0--4000 {cm$^{-1}$}.
   The calculations are performed using the Multiconfiguration
   Time-Dependent Hartree (MCTDH) method for propagation of
   wavepackets.
   All the fundamentals and several overtones of the vibrational
   motion are computed. 
   The spectrum of H$_5$O$_2^+$ is shaped to a large extent by
   couplings of the proton-transfer motion to large amplitude fluxional
   motions of the water molecules, water bending and water-water
   stretch motions. These couplings are identified and discussed, and the
   corresponding spectral lines assigned.

   The large couplings featured by H$_5$O$_2^+$ do not hinder, however, to describe
   the coupled vibrational motion by well defined simple types of vibration
   (stretching, bending, etc.) based on well defined modes of vibration, in terms
   of which the spectral lines are assigned.
   Comparison of our results to recent
   experiments and calculations on the system is given.
   The reported MCTDH IR-spectrum is in very good agreement to the recently measured
   spectrum by Hammer {\em et al.} \mbox{[JCP, \textbf{122}, 244301, (2005)]}.
\end{abstract}


\maketitle

\section{Introduction}


The understanding of the hydrated proton in bulk water and water-containing systems 
is of importance to many areas in chemistry and biology.
Much effort has been directed in recent years towards obtaining a better understanding
of the excess proton in water and important
contributions have appeared due to the advance in experimental and theoretical 
techniques \cite{mar99:601,agm99:493,jia00:1398,hea05:1765,ham05:244301}. 

The importance of the hydrated proton and the amount of work devoted to the problem
contrast with the fact that the smallest system in which a proton is shared between
water molecules, the protonated water dimer ({\zun}), is not yet completely understood,
and an explanation of some important spectral signatures and an overall picture
of the dynamics of the cation is lacking.
Accurate infrared (IR) spectroscopy of protonated
water clusters prepared in the gas phase has become possible in recent
years \cite{asm03:1375,fri04:9008,hea04:11523,hea05:1765,ham05:244301}, opening the
door to a deeper understanding of these systems and the hydrated proton in general.

%
%
The protonated water dimer, {\zun}, also known as Zundel cation, is 
the smallest protonated water
cluster, and has been recently object of intense study.  The infrared
(IR) spectrum of the system has been measured in the gas phase,
either using multiphoton dissociation techniques
\cite{asm03:1375,fri04:9008} or measuring the vibrational
predissociation spectrum of {\zun}$\cdot$RG$_n$ clusters with
RG=Ar,Ne \cite{hea04:11523,ham05:244301}. The obtained spectra
could not be consistently assigned in terms of fundamental
frequencies and overtones of harmonic vibrational modes due to
large-amplitude anharmonic displacements and couplings of the
cluster. Hence, more sophisticated theoretical approaches are
required. Several theoretical studies have been conducted over
the last years in order to understand and assign the IR spectrum
of the  cation
\cite{ven01:240,dai03:6571,hua05:044308,ham05:244301,sau05:1706,kal06:2933,ven07:letter,ven07:paperI}.

%
%
 The first measurement \cite{asm03:1375} of the IR multiphoton
 dissociation spectrum (IRMPD) of {\zun} spanned the range
 between 620 and 1900 {\icm}.  Three main absorptions were
 discussed and assigned, based on a previous quantum-dynamical
 simulation of the IR absorption spectrum on a 4D model of the
 hydrogen-bond (O-H-O) fragment \cite{ven01:240}.  Those
 assignments were revisited in the context of newer IRMPD
 experiments and calculations, producing somewhat disparate
 results \cite{fri04:9008,ham05:244301,sau05:1706,kal06:2933}.
%
%
Recent measurements of the IR predissociation spectrum of the {\zun} cation
in argon-solvate \cite{hea04:11523} and neon- and
argon-solvate \cite{ham05:244301} conditions present spectra with a simpler structure
than the multiphoton IRMPD ones.
It is expected that the spectrum of the {\zun}$\cdot$Ne$_1$ complex is close
to the linear absorption spectrum of the bare cation \cite{ham05:244301}.
This spectrum features a doublet structure in the region of 1000 {\icm}
made of two well-defined absorptions at
928 {\icm} and 1047 {\icm}.
This doublet structure was not fully understood until recently, although
the highest-energy component had been already assigned by Bowman and collaborators to the
asymmetric proton-stretch fundamental (\asymH) \cite{ham05:244301}.
A similar argument was made by Sauer and collaborators based on 
classical-trajectories calculations \cite{sau05:1706}.
In this respect, it is known from recent classical-dynamics simulations on accurate potential
energy surfaces (PES) that the {\asymH} motion features large
amplitude displacements strongly coupled to other modes of the system. The
central-proton displacement would then be involved in most of the lines appearing in
the IR spectrum, since this motion contributes the largest changes in the dipole
moment of the cation \cite{sau05:1706,kal06:2933}.
Recent work by us assigns unambiguously the doublet for the first time \cite{ven07:letter}, which is
shown to arise from the coupling of low-frequency motions of the water 
molecules to the central proton motion.

%
%
In this work we undertake the simulation of the IR spectrum and dynamics of {\zun} using
the multiconfiguration time-dependent Hartree (MCTDH) \cite{man92:3199,bec00:1,mey03:251} 
method. In doing so we do not rely on a
low-dimensional model of the system, but we treat it in its full (15D) 
dimensionality. The obtained IR spectrum is compared to recent
experimental and theoretical results \cite{ham05:244301}. The experimental
IR spectrum and our results agree well in the position and relative intensities 
of the main spectral features, which allows us to extract meaningful 
conclusions regarding the dynamics of the cluster.
All fundamentals and several overtones of the vibrational motion 
of the cation are
computed and their properties analyzed. 
The computation of vibrational states in terms of their fully correlated wavefunctions
is invaluable in the assignment of spectral lines arising from the time-dependent
computation of the IR spectrum.

The reported simulations are performed using curvilinear coordinates, which
has been found crucial for the description of the large amplitude motions 
and anharmonicities featured by the cluster \cite{ven07:paperI}. 
A description of the derivation of the exact kinetic energy operator (KEO) 
in this set of coordinates and of the representation of the potential
energy surface (PES) are discussed in the companion paper \cite{ven07:paperI}, 
hereafter referred to as Paper I. The quality of the PES expansion,
and the properties of the ground-vibrational
state of the system are also analyzed in Paper I.
The reference PES and dipole moment surfaces (DMS) used are those of Huang et al.
\cite{hua05:044308}, which constitute the most accurate {\em ab initio} surfaces
available to date for this system.

\section{Theory and Methods}   \label{sec:theory}

 The quantum-dynamical problem is solved using
 the multiconfiguration time-dependent Hartree (MCTDH) method \cite{man92:3199,bec00:1,mey03:251}.
 For a brief description of MCTDH see Paper I. 
 All the reported simulations
 were performed with the Heidelberg MCTDH package of programs
 \cite{mctdh:package}.

 The definition of the system of curvilinear coordinates used in the description of {\zun}
 is given in Paper I. The 15 internal coordinates are briefly reintroduced here to make
 the discussion self-contained. These are:
 the distance between the centers of mass of both water molecules ($R$),
 the position of the central proton 
 with respect to the center of mass of the water dimer ($x$,$y$,$z$),
 the Euler angles defining the relative orientation between the two water
 molecules (waggings: $\gamma_A,\gamma_B$; rockings: $\beta_A,\beta_B$;
 internal relative rotation: $\alpha$) and the Jacobi coordinates
 which account for the particular configuration of each water 
 molecule ($R_{1(A,B)},R_{2(A,B)},\theta_{(A,B)})$) where $R_{1x}$ is the
 distance between the oxygen atom and the center of mass of the
 corresponding H$_2$ fragment, $R_{2x}$ is the H--H distance and
 $\theta_{x}$ is the angle between these two vectors.
 The grouping of the coordinates into modes is discussed in Paper I. Here we repeat
 only the mode-combination scheme used. The coordinates have been grouped
 into the following five combined modes:
  $Q_1 = [z,\alpha,x,y]$,
  $Q_2 = [\gamma_A,\gamma_B]$,
  $Q_3 = [R, u_{\beta_{A}}, u_{\beta_{B}}]$,
  $Q_4 = [R_{1 A}, R_{2 A}, u_{\theta_{1 A}}]$ and
  $Q_5 = [R_{1 B}, R_{2 B}, u_{\theta_{1 B}}]$.

  \subsection{IR Spectrum calculation in the MCTDH framework}

  The IR absorption cross-section is given by \cite{bal90:1741}:
  \begin{equation}
    \label{eq:spectrum-ti}
     I(E) = \frac{\pi E}{3\,c\,\epsilon_0\,\hbar} 
            \sum_n |\langle \Psi_{n}| \Psi_{\mu,0} \rangle|^2 \; 
     \delta(E+E_0 - E_n),
  \end{equation}
  where $|\Psi_{\mu,0}\rangle$ is the dipole-operated initial state, i.e.,
  $|\Psi_{\mu,0}\rangle \equiv \hat{\mu} \, |\Psi_0 \rangle$, and $E_0$ is the ground-state energy.
  The IR spectrum may be equivalently computed in the time-dependent picture.
  From Eq. (\ref{eq:spectrum-ti}) follows
  \begin{subequations}
    \label{eq:spectrum-td}
   \begin{eqnarray}
     I(E) & = & \frac{E}{6\,c\,\epsilon_0\,\hbar^2} 
                \int_{-\infty}^{\infty} e^{i(E+E_0 - E_n)t/\hbar}
                       \langle \Psi_{\mu,0} | \Psi_{n} \rangle \; 
                       \langle \Psi_{n} | \Psi_{\mu,0} \rangle \;dt  \\
          & = & \frac{E}{6\,c\,\epsilon_0\,\hbar^2} 
                \int_{-\infty}^{\infty} e^{i(E+E_0)t/\hbar}
                       \langle \Psi_{\mu,0} | e^{-i\;H\;t/\hbar} | \Psi_{\mu,0} \rangle \;dt \\
          & = & \frac{E}{3\,c\,\epsilon_0\,\hbar^2}\, 
                Re\int_{0}^{\infty} e^{i(E+E_0)t/\hbar}
                       a_\mu(t) \;dt,
   \end{eqnarray}
  \end{subequations}
  where the autocorrelation function of the dipole-operated initial state, $a_\mu(t)$, has
  been defined in Eq. (\ref{eq:spectrum-td}c) implicitly.
  The real part, $Re$, appears because $a_\mu(-t)=a_\mu^{\ast}(t)$ holds.
  Eq. (\ref{eq:spectrum-td}c) tells that the IR spectrum is obtained in the time-dependent picture by 
  Fourier transformation (FT) of the autocorrelation of the dipole-operated initial state.
  Whether a time-dependent or a time-independent approach is more efficient depends on the
  problem at hand.
  A time-independent approach relies on the accurate computation of the eigenstates
  of the Hamiltonian, $|\Psi_{n}\rangle$, which may be accomplished by iterative-diagonalization methods
  of the Hamiltonian matrix expressed in some basis, obtaining all the eigenstates up to some
  desired energy, and then using Eq. (\ref{eq:spectrum-ti}) to compute the IR spectrum.
  The method to obtain eigenstates of the Hamiltonian within the MCTDH approach is called 
  {\em improved relaxation} \cite{mey06:179} and consists essentially 
  on a time-independent multiconfiguration self consistent field (MCSCF) computation of an eigenstate, in which 
  both the expansion coefficients of the wavefunction and the basis of SPFs (see Eq. (1) in Paper I) 
  are optimized to self-consistency.
  The optimal basis of SPFs varies from
  state to state, which difficults the application of iterative diagonalization procedures to the 
  computation of the whole spectrum. To obtain each excited state with {\em improved relaxation},
  a new computation is needed in which the basis is optimized for that specific state.
  A comprehensive calculation of an IR spectrum by {\em improved relaxation} has been recently achieved 
  for 6D problems, namely H$_2$CS \cite{mey06:179}, and HONO \cite{ric07:xxx}. 
  The lowest 184 states of $A^\prime$ symmetry
  of the latter molecule were calculated.
  The {\em improved relaxation} approach, however, is not practicable with the current computational
  capabilities to obtain the whole spectrum for a molecule of the size of {\zun}.
  The iterative process of improved relaxation converges only if the space spanned by the
  configurations allows to separate the desired state from neighboring ones. When the
  density of states increases, the number of SPFs necessary to ensure convergence increases
  as well. This makes it impossible to compute states higher in energy. For the present problem
  we could compute states up to about 600 {\icm}. However, not only the excitation energy but also
  state coupling determines the feasibility of improved relaxation. For example, we could compute
  the second excited state of the water-water stretch by {\em improved relaxation}, 
  despite its excitation energy is {\ESRB} {\icm}.

  Propagation of a wavepacket by MCTDH, on the other hand, is always feasible. A
  small set of SPFs makes a propagation less accurate, but not impossible.
  Therefore, the full IR spectrum is efficiently calculated
  in the time-dependent representation.
  The resolution at which different peaks of the spectrum are resolved
  is given by the Fourier Transform (FT) of the damping function $g(t)$ \cite{bec00:1} with which
  the autocorrelation $a_\mu(t)$ is to be multiplied when performing the
  integral (\ref{eq:spectrum-td}c)
  to minimize artifacts due to the Gibbs phenomenon.
  We choose $g(t)=\cos(\pi t/2T)$ and set
  $g(t)=0$ for $t>T$, where $T$ denotes the length of the autocorrelation
  function. Since we make use of the $T/2$ trick, 
  \mbox{$a_\mu(t) = \langle\Psi_{\mu 0}^{\ast}(t/2)|\Psi_{\mu 0}(t/2)\rangle$}, which
  holds when the initial wavepacket is real and the Hamiltonian symmetric,
  $T$ is twice the propagation time. The FT of $g(t)$ is known \cite{bec00:1} and its full
  width at half maximum (FWHM) is
  \begin{equation}
    \label{eq:resolution}
      \Delta E = 27,300 \; \textrm{cm}^{-1} \; \textrm{fs} \; / T.
  \end{equation} 
  To clearly identify two peaks 100 {\icm} apart, i.e., the doublet structure in the
  {\zun} spectrum, one needs a resolution of $\Delta E=67$ {\icm} and hence a propagation
  time of $T/2=200$ fs. Time propagations have been carried out over 500 fs providing a resolution
  of 30 {\icm}.

  Often we make use of the filter diagonalization (FD) technique \cite{wal95:8011,bec98:3730,bec01:2036,gat01:1576} to
  analyze the autocorrelation function of the propagated wavepackets, which
  yields eigenenergies and spectral intensities.
  FD is able to provide accurate eigenenergies
  from shorter time propagations than the ones needed to resolve the IR spectrum.
  However, FD calculations are more sensitive to errors in the autocorrelation function compared
  to the FT method. 
  Therefore these calculations
  are targeted to obtain accurate energies of a single state or a small group of related states from
  a propagation of an appropriately prepared initial wavepacket.

  The dipole operator is a vector and the operator $\hat{\mu}$ appearing in 
  Eqs. (\ref{eq:spectrum-ti},\ref{eq:spectrum-td}) is to be interpreted as
  the scalar product $\vec{e}\cdot\vec{\hat{\mu}}$ where $\vec{e}$ denotes the polarization
  vector of the absorbed light. As the molecule is assumed to be randomly oriented
  we average over all orientations and obtain
  \begin{equation}
    \label{eq:intensities}
      I(E)=\frac{1}{3}(I_z(E)+2 I_x(E)),
  \end{equation}
  where $I_z$ and $I_x$ denote the intensities obtained with the $x-$ and $z-$component
  of the dipole operator, respectively. The factor 2 appears because $a_{\mu_x}=a_{\mu_y}$ holds.
  The intensity $I_y$ is hence not explicitly computed.

  \subsection{Eigenstates of the system and spectrum assignment}  \label{sec:relaxation}

  Even if the computation of the IR spectrum in the time-dependent representation is feasible,
  a means of assigning the different lines to specific motions of the system is still required.

  Wavefunctions of excited states converged by {\em improved relaxation} \cite{mey06:179} contain all the possible
  information of that specific state. 
  The intensity of a given excited state $|\Psi_{n}\rangle$ is readily obtained by computing
  the dipole moment
  $|\langle\Psi_{n}|\hat{\mu}|\Psi_{0}\rangle|^2$. 
  But even if an excited state of interest $|\Psi_{n}\rangle$ has been obtained it is difficult 
  to directly inspect these mathematical objects due to their high dimensionality.
  Moreover, for the higher excited states we do not have $|\Psi_{n}\rangle$ at our disposal
  but only an autocorrelation function providing spectral lines. In both cases we
  characterize the eigenstates by their overlaps with carefully chosen test states,
  i.e., by the numbers $|\langle\Phi_{test}|\Psi_n\rangle|^2$.
  The following procedures are used:
       \begin{itemize}
       \item[1.]
       Test states $|\Phi_{test}\rangle$ are generated
          \begin{itemize}
             \item[a)]
              by applying some operator $\hat{O}$ to a previously
              converged eigenfunction,
              \begin{equation}
                \label{eq:test1}
                   |\Phi_{test}\rangle=N\hat{O}|\Psi_{n}\rangle,
              \end{equation}
              where $N$ is a normalization constant, e.g.,
              $N\hat{z}|\Psi_{0}\rangle$ generates a test state which 
              in essence differs from the ground state $|\Psi_{0}\rangle$
              by a one quantum excitation in the proton-transfer coordinate $z$.
             \item[b)]
              by forming Hartree products, where the SPFs are obtained 
              through diagonalization of mode-Hamiltonians $\hat{h}_j$.
              The $\hat{h}_j$ are low-dimensional Hamiltonians and each $\hat{h}_j$
              operates on the space of a group of coordinates (see Paper I
              for details on coordinates grouping into modes).
              Rather than using single Hartree products one may use
              linear combinations of products in order to satisfy a symmetry constraint.
          \end{itemize}
       \item[2.]
       The overlaps
       $|\langle\Phi_{test}|\Psi_{n}\rangle|^2$ are then computed by
          \begin{itemize}
             \item[a)]
              by direct evaluation of the scalar product if $|\Psi_{n}\rangle$ is available.
             \item[b)]
              by Fourier transform of the autocorrelation function
              \mbox{$a(t)=\langle\Phi_{test}|\exp(-i\;H\;t)|\Phi_{test}\rangle$}.
              The overlap is obtained via the formula \cite{mey06:179}:
              \begin{equation}
                \label{eq:int2}
                  |\langle\Phi_{test}|\Psi_{n}\rangle|^2=\frac{\pi}{2T}Re 
                  \int_0^{T} e^{iE_nt} a(t) \cos(\frac{\pi t}{2T}) dt
              \end{equation}
             \item[c)]
              by Fourier transformation of the cross-correlation function
              \mbox{$c(t)=\langle\Phi_{test}|\exp(-i\;H\;t)|\Psi_{\mu,0}\rangle$}.
              The absolute square of the FT of $c(t)$ at energy $E_n$ must then be divided by the
              spectral intensities of $|\Psi_{\mu,0}\rangle$ to obtain $|\langle\Phi_{test}|\Psi_{n}\rangle|^2$.
              This is the fastest method because it does not require additional propagations.
              However, it is also the least accurate procedure. We used this method
              several times to obtain a quick overview, but it was not used to generate
              data reported in this article.
          \end{itemize}
       \end{itemize}

    Excitations related to test states will be denoted $(n_1\,q_1,n_2\,q_2\ldots)$, where $n_1$ represents
    the quanta of excitation on coordinate $q_1$ (in a separable limit). 
    A test state will be denoted by $|\Phi_{n_1\,q_1,n_2\,q_2\ldots}\rangle$
    where the terms with $n_j=0$ are omitted.
    In case a test state has been generated such that it cannot be represented by this simple notation,
    it will be defined more explicitly.

\section{Results and Discussion}

\begin{figure}[h!]
      \begin{center}
        \includegraphics[width=8.5cm]{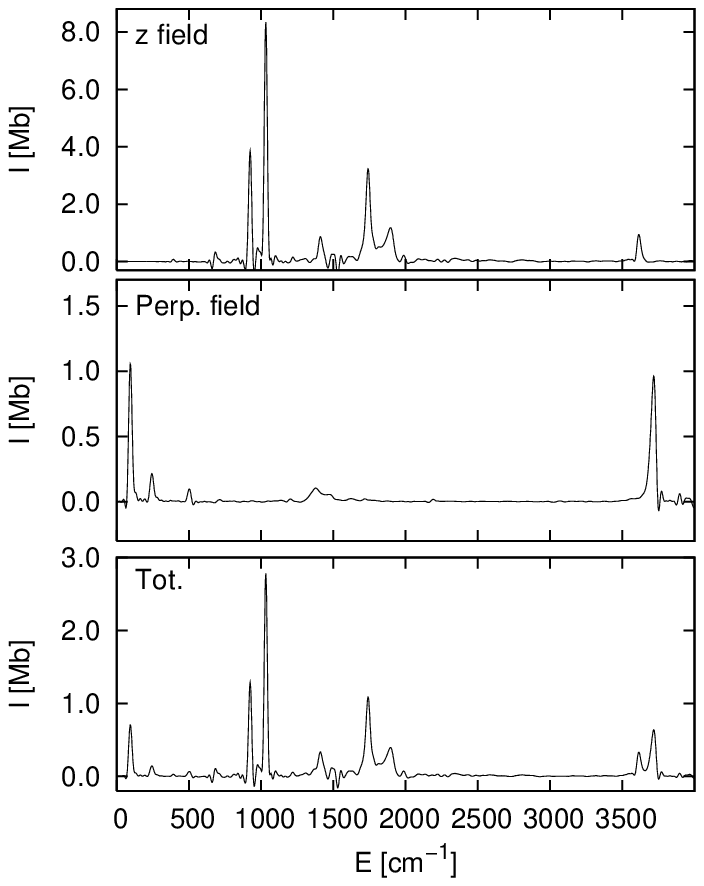}
      \end{center}
     \caption{\figfoot}
     \label{fig:IR04000}
    \end{figure}

    \placefig{fig:IR04000}

    \placefig{fig:IRparts}

    The MCTDH spectrum in the full range 0--4000 {\icm} is depicted in Fig.~\ref{fig:IR04000}
    and compared to experiment in Fig.~\ref{fig:IRparts}.
    The dipole-moment operated ground state $\hat{\mu}|\Psi_{0}\rangle$ was propagated for
    500 fs, yielding an autocorrelation of 1000 fs. The spectrum
    was calculated according to Eq. (\ref{eq:spectrum-td}) and the FWHM resolution of the spectrum is,
    according to Eq. (\ref{eq:resolution}), about 30 {\icm}.

    In the following sections the different parts of the spectrum and the peak assignments are discussed.
    Table \ref{tab:states} collects
    the energies of several states of {\zun} obtained with MCTDH, MULTIMODE \cite{ham05:244301} 
    and experimental values on the {\zun$\cdot$Ne} system \cite{ham05:244301}.
    \placetab{tab:states}
    The symmetry labels of the discussed vibrational states are given
    within the $\mathcal{G}_{16}$ symmetry group, which arises due to the feasibility
    (in the sense of Longet-Higgins \cite{lon63:445}) of the wagging and internal rotation motions (see Paper I). 
    A character table for the $\mathcal{G}_{16}$ symmetry group is found in Table A-25 in Ref. \cite{bunker-book:symmetry}.
    If the internal rotation motion around $\alpha$ is assumed to be unfeasible, 
    the symmetry analysis can be performed using the $\mathcal{D}_{2d}$ point group.
    The permutation-inversion group $\mathcal{G}_{16}$
    contains the $\mathcal{D}_{2d}$ point group as a subgroup,
    but allows additionally to permute the two hydrogens of one of the
    water monomers \cite{wal99:10403,bunker-book:symmetry}.
    The $\mathcal{G}_{16}$ labels
    reduce to the $\mathcal{D}_{2d}$ ones by ignoring ($-/+$) signs in the labeling provided by $\mathcal{G}_{16}$.
    The use of $\mathcal{G}_{16}$ becomes important when labeling the states related to the internal rotational
    motion. The $z$ component of the dipole moment excites vibrational states of $B_2^+$ symmetry, 
    while the perpendicular components those of $E^+$ symmetry.
    
    \subsection{Low-energy region}

  The low energy region of the spectrum, below 900 {\icm}, has not
  yet been accessed experimentally. The reported spectrum in Fig.~\ref{fig:IR04000}
  shows a strong absorption around 100 {\icm} followed by two lines of fewer intensity
  at about 250 and 500 {\icm}, respectively. The three absorptions arise from excitation
  of the perpendicular component of the field.
  Modes oscillating at these low frequencies are strongly anharmonic: 
  already in the ground vibrational state, the system
  interconverts between equivalent
  minima through low barriers along the wagging ($\gamma_A$, $\gamma_B$) and torsional
  ($\alpha$) coordinates \cite{ven07:paperI}. 

  \placefig{fig:dens3}
  The first three excited states associated to the internal
  rotation ($1\alpha$, $2\alpha$, $3\alpha$) have been computed 
  with {\em improved relaxation} and their
  probability-density along the internal rotation $\alpha$ is
  depicted in Fig.~\ref{fig:dens3}. The energy of the (4$\alpha$) state
  has been computed by FD.
  They have excitation energies of {\ESrotO}, {\ESrotA}, {\ESrotB} and {\ESrotC} {\icm},
  and symmetries $A_1^-$, $B_1^+$, $B_1^-$ and $A_1^+$, respectively.
  Here ($1\alpha$) is the splitting state related to the torsional barrier along $\alpha$.
  The excitation energy of 1 {\icm} is hence the tunneling split.
  In fact, the symmetry label of the rotationally-split state of a $(+)$ state
  is obtained by multiplication with $A_1^{-}$.

  \placefig{fig:dens1}
  \placefig{fig:dens2}
  Fig.~\ref{fig:dens1} depicts the probability-density 
  projection on the wagging 
  coordinates for the ground vibrational state, which is of $A_1^+$ symmetry, as well as
  for one of the two fundamental states ($w_{1a}$,$w_{1b}$) of the wagging
  modes. The fundamental wagging modes are centered at {\ESwagA} {\icm} and 
  belong to the $E^-$ irreducible representation of
  the $\mathcal{G}_{16}$ group. Therefore, they are doubly degenerate, dark states.
  The band at about 100 {\icm} (see Fig.~\ref{fig:IR04000} center) due to the perpendicular 
  component of the dipole corresponds to the states centered at {\ESwagAaa} {\icm}, which 
  belong to the $E^+$ irreducible representation. Such states are combined states of
  the fundamental wagging motion and the ($1\alpha$) state of the internal rotation, of
  $A_1^-$ symmetry. Thus, both pairs of degenerate wagging states, 
  centered at {\ESwagA} and {\ESwagAaa} {\icm} are
  split states with respect to the internal rotational barrier around $\alpha$.
  In a similar way the band at about 250 {\icm} corresponds to the combination of
  the fundamental wagging motion and the ($3\alpha$) internal-rotation state, which
  belongs to the $B_1^-$ irreducible representation. This combination results in two
  degenerate states of $E^+$ symmetry.
  The energies of the next three wagging-mode states ($w_{2}$,$w_{3}$,$w_{4}$)
  are, respectively, {\ESwagB}, {\ESwagC} and {\ESwagD} {\icm} and they are shown in 
  Figs. \ref{fig:dens2}a, \ref{fig:dens2}b and \ref{fig:dens2}c, respectively.
  These three states correspond to two quanta of excitation in the wagging motions
  and they can be represented by kets
  $|11\rangle$, \mbox{$(|20\rangle-|02\rangle)/\sqrt{2}$}
  and \mbox{$(|20\rangle+|02\rangle)/\sqrt{2}$}, respectively, where the $|ab\rangle$ notation
  signifies the quanta of excitation in the wagging motions of monomer $A$ and $B$.
  These states have symmetries $B_1^+$, $B_2^+$ and $A_1^+$, respectively.
  In the harmonic limit these three states would be degenerate. The next 2 states, $w_{4a}$,$w_{4b}$,
  which are not computed, are degenerate again and correspond to kets
  \mbox{$|31\rangle$} and \mbox{$|13\rangle$}.
  State $w_{2}$ has an energy
  that nearly doubles the energy of the $w_{1x}$ states, since it
  roughly corresponds to one quantum in state $w_{1a}$ and
  one quantum in state $w_{1b}$. The strong anharmonicity of the wagging
  motions as well as the coupling between right and left wagging
  can be further appreciated in the progression of $w_{2}$, $w_{3}$ and
  $w_{4}$ vibrational-state energies.
  We emphasize again that these three states are degenerate in the harmonic limit.
  In addition, the harmonic-analysis energies
  of the two lowest wagging-fundamentals $w_{1a}$ and $w_{1b}$ are around
  300 {\icm} larger than the MCTDH result and do not account for their
  degeneracy, since harmonic normal-modes are constructed taking as a reference
  the $\mathcal{C}_{2}$ absolute minimum. However, as discussed above and in Paper I,
  {\zun} interconverts through low potential barriers between several equivalent minima
  and has $\mathcal{G}_{16}$ symmetry.
  The state $w_{3}$ has four probability-density maxima along the 2D space spanned
  by $\gamma_A$ and $\gamma_B$. They correspond to geometries in which one of the
  water molecules adopts a trigonal-planar geometry (H$_3$O$^+$ character) and the
  other adopts a pyramidal geometry (H$_2$O character). 
  This state transforms according to the $B_2^+$ symmetry representation, which
  is also the symmetry of the proton-transfer fundamental.
  State $w_{3}$ will play a major
  role due to its strong coupling to the proton-transfer mode.

  The first two fundamentals of the symmetric stretch ({\symOO}, $R$ coordinate) have 
  energies of {\ESRA} and {\ESRB} {\icm} respectively, and have $A_1^+$ symmetry,
  while the rocking fundamentals, which are degenerate $E^+$ states, have an energy of {\ESrock} {\icm}
  and are responsible of the band appearing slightly below 500 {\icm} in the MCTDH spectrum.
  In contrast to the wagging motion, the rocking motion is fairly harmonic and exhibits
  only a weak coupling between left and right rocking. The energies of the two-quanta
  rocking states $r_2$, $r_3$ and $r_4$ -- defined in a similar way to $w_{2}$, $w_{3}$ and $w_{4}$
  but with a reverse ordering of the zero-order {\em ket}-states --
  are therefore almost degenerate. These energies read {\ESrockD}, {\ESrockC} and {\ESrockB} {\icm}.

  \subsection{Doublet at 1000 {\icm}}

\begin{figure}[h!]
      \begin{center}
        \includegraphics[width=8.5cm]{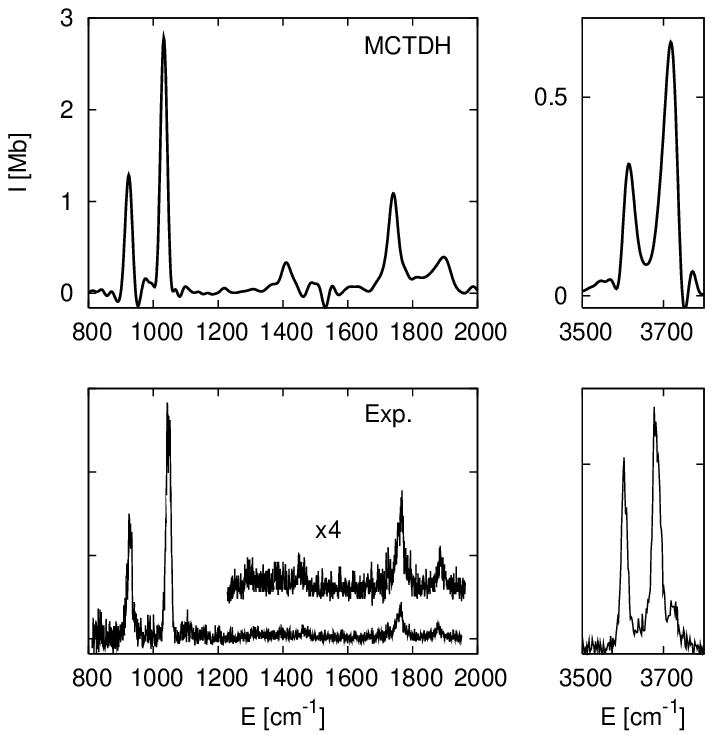}
      \end{center}
     \caption{\figfoot}
     \label{fig:IRparts}
    \end{figure}

\begin{figure}[h!]
      \begin{center}
        \includegraphics[width=8.5cm]{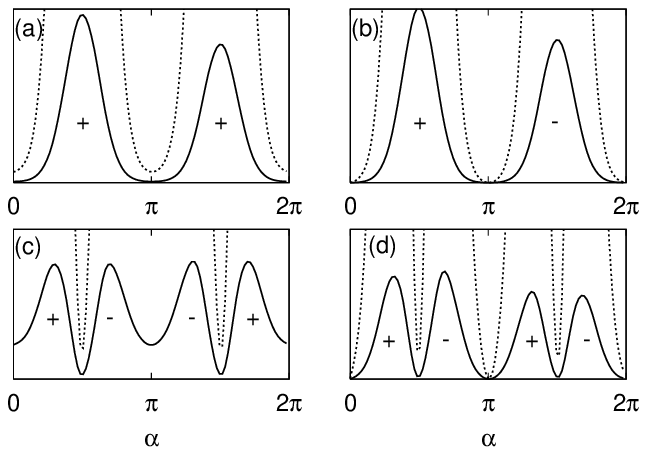}
      \end{center}
     \caption{\figfoot}
     \label{fig:dens3}
    \end{figure}

\begin{figure}[h!]
      \begin{center}
        \includegraphics[width=8.5cm]{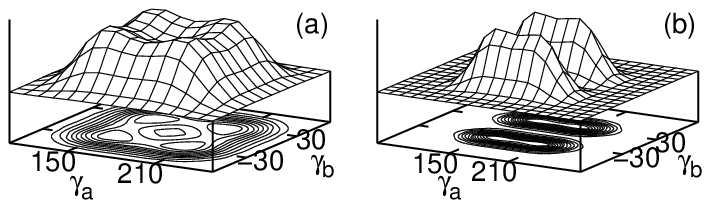}
      \end{center}
     \caption{\figfoot}
     \label{fig:dens1}
    \end{figure}

\begin{figure}[h!]
      \begin{center}
        \includegraphics[width=12.75cm]{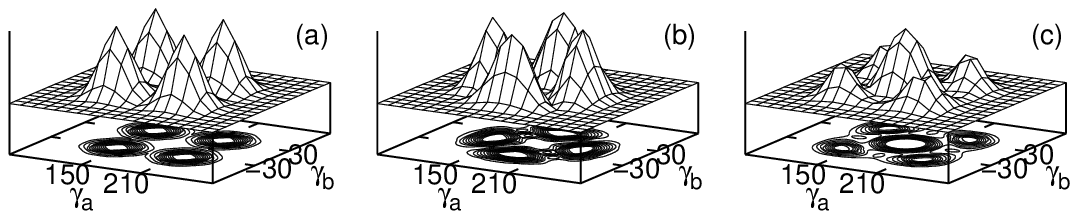}
      \end{center}
     \caption{\figfoot}
     \label{fig:dens2}
    \end{figure}

\begin{figure}[h!]
      \begin{center}
        \includegraphics[width=8.5cm]{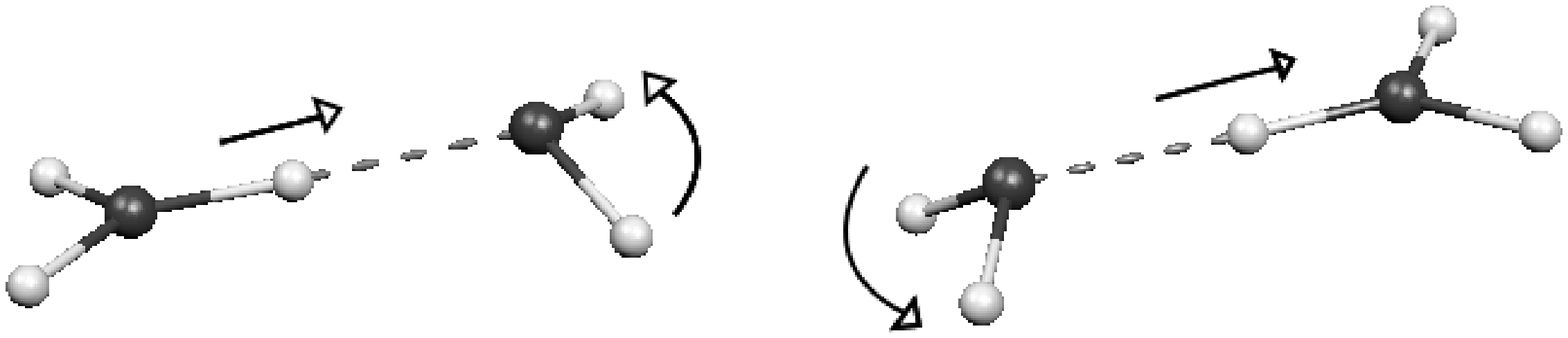}
      \end{center}
     \caption{\figfoot}
     \label{fig:pictorial}
    \end{figure}

  The doublet centered at 1000 {\icm} is the most characteristic feature of the
  IR spectrum of {\zun}. It is depicted in Fig.~\ref{fig:IRparts} (top). The highest
  energy line has been measured to be at 1047 {\icm} while the low energy component
  appears at 928 {\icm} \cite{ham05:244301}. 
  There is accumulated evidence in the literature that the absorption of the proton-transfer fundamental occurs in the
  region of 1000 {\icm} \cite{dai03:6571,hua05:044308,ham05:244301,sau05:1706,kal06:2933}.
  Specifically, the band at 1047 {\icm} in Ref. \cite{ham05:244301} was assigned
  to the first excitation of the central proton motion \cite{ham05:244301} based on 
  MULTIMODE \cite{bow03:533} calculations. This band is the most intense band of the
  spectrum since the central proton motion 
  along the $z$ axis induces a large change in the dipole-moment of the cation.
  The low-energy component has been recently assigned by us \cite{ven07:letter}. The doublet
  is seen to arise from coupling between the proton-transfer motion {\asymH}, the 
  low frequency water-wagging modes and the water-water stretching {\symOO} motion.
 In order to obtain a fundamental understanding
 of the low-energy ($|\Psi_d^l\rangle$) and high energy ($|\Psi_d^h\rangle$)
 components of the doublet, test states were constructed
 by operating with $\hat{z}$ on the ground state: $|\Phi_{1z}\rangle = \hat{z}|\Psi_0\rangle N$,
 where $N$ is a normalization constant,
 and by operating with $(\hat{R}-R_0)$ on the third excited wagging state $w_3$:
 $|\Phi_{1R,w_3}\rangle = (\hat{R}-R_0)|\Psi_{w_3}\rangle N$.
 Note that $| \Phi_{1z}\rangle$ is characterized by one quantum of excitation in the
 proton-transfer coordinate whereas $| \Phi_{R1,w_3}\rangle$ by one quantum in
 {\symOO} and two quanta in the wagging motion.
 These two test states were propagated and their auto- and crosscorrelation
 functions were used for FD analysis, which yielded an energy of {\ESdoubl} {\icm}
 for $|\Psi_d^l\rangle$ and an energy of {\ESdoubh} {\icm} for $|\Psi_d^h\rangle$.
 These energies are in good accordance to the peaks in Fig.~\ref{fig:IRparts} which
 arise from the propagation of $|\Psi_{\mu,0}\rangle$.
 The spectral intensities were also obtained by FD analysis.
 The overlaps of the test states to the states making the doublet read:
 $|\langle \Phi_{1z} | \Psi_d^l \rangle|^2 = 0.09$,
 $|\langle \Phi_{1R,w_3} | \Psi_d^l \rangle|^2 = 0.83$ and
 $|\langle \Phi_{1z} | \Psi_d^h \rangle|^2 = 0.46$,
 $|\langle \Phi_{1R,w_3} | \Psi_d^h \rangle|^2 = 0.10$.
 One should take into account that these numbers depend on the exact
 definition of the test states, which is not unique.
 However, they provide a clear picture of the nature of the doublet:
 the low-energy band has the largest contribution from the combination of the
 symmetric stretch and the third excited wagging (see Figs. \ref{fig:dens2}b and
 \ref{fig:dens3}c), whereas
 the second largest is the proton-transfer motion. For the high-energy band
 the importance of these two contributions is reversed.
 Thus, the doublet may be regarded as a Fermi-resonance between two zero-order
 states which are characterized by ($1R$, $w_3$) and ($1z$) excitations, respectively.
 The reason why the third wagging excitation
 plays an important role in the proton-transfer doublet is
 understood by inspecting Fig.~\ref{fig:dens2}b and \ref{fig:pictorial}.
 The probability density of state $w_3$ has four maxima, each of which
 corresponds to a planar conformation of H$_2$O-H$^+$ (H$_3$O$^+$ character)
 for one of the waters, and a bent conformation (H$_2$O character) where
 a lone-pair H$_2$O orbital forms a hydrogen bond with the central proton. When the proton
 oscillates between the two waters, the two conformations exchange their
 characters accordingly.
 Thus, the asymmetric wagging mode ($w_3$, {\ESwagC} {\icm}) combines with
 the water-water stretch motion ($R$, {\ESRA} {\icm}) to reach an energy close to the
 natural absorption-frequency of the proton transfer, making these motions coupled.
 The two states of the doublet transform according to the $B_2^+$ irreducible representation
 of $\mathcal{G}_{16}$.

 One last remark regarding the spectral region of the doublet is the small but noticeable
 absorption, which is appreciated both in the experimental and MCTDH spectra between both
 peaks (see Fig.~\ref{fig:IRparts}). This low-intensity absorption is due to the $r_3$ rocking
 centered at {\ESrockC} {\icm} and which belongs to the $B_2^+$ irreducible representation.
 In an analogous way to the $w_3$ wagging state, the $r_3$ rocking state presents four 
 probability-density maxima along the rocking coordinates, each of which consists
 of one water aligned with the central axis while the other is in a bent conformation, and
 therefore will present some degree of coupling to the proton-transfer motion.
 The low absorption of this band, despite its proximity to the natural absorption-frequency
 of the proton-transfer motion, is qualitatively explained by the fact that the rocking
 motions do not change the hybridization properties of the water monomers and are 
 of a lower amplitude in comparison to the wagging motions. The rocking motion hence couples
 more weakly to the proton-transfer motion and induces a smaller change of the dipole-moment
 as compared to the wagging motions.

 \placefig{fig:pictorial}
 
    \subsection{1000-2000 {\icm} region}

       The region between the proton-transfer doublet and the doublet centered at
       1800 {\icm} features couplings related to the {\asymH} and {\symOO} motions.
       The MCTDH spectrum reported in Fig.~\ref{fig:IR04000}
       presents three main absorptions in this range, located at {\ESzARA}, {\ESbendOP} and
       {\ESzARB} {\icm}. We call the eigenstates producing these peaks $|\Psi_{m1}\rangle$, $|\Psi_{m2}\rangle$
       and $|\Psi_{m3}\rangle$, respectively, where the $m$ stands for 
       middle spectral-range.
       The experimental {\zun}$\cdot$Ne spectrum shows two clearly
       distinguishable bands at similar positions to the {\ESbendOP} and
       {\ESzARB} {\icm} absorptions in the MCTDH spectrum, as depicted in Fig.~\ref{fig:IRparts}.
       The spectrum of {\zun}$\cdot$Ne also shows weak but non-negligible absorption in the
       region immediately above 1400 {\icm} (see Fig. 5 in Ref. \onlinecite{ham05:244301}).
       
       Propagation of test states followed by Fourier analysis of their autocorrelation
       functions as described in 
       Sec. \ref{sec:relaxation} were used to assign these peaks. The following test
       states were generated using eigenfunctions of low dimensional Hamiltonians: 
       $|\Phi_{1z,1R}\rangle$, $|\Phi_{1z,2R}\rangle$, $|\Phi_{bu}\rangle$.
       The test state $|\Phi_{bu}\rangle$ consists on the water-bending with {\em ungerade} symmetry,
       i.e. it is characterized by $(|01\rangle-|10\rangle)/\sqrt{2}$ where the two entries
       indicate the quanta of bending motion of monomer A and B, respectively.
       \placetab{tab:matrix}
       
       The overlaps $|\langle\Phi|\Psi\rangle|^2$, where $|\Phi\rangle$ is a test state and
       $|\Psi\rangle$ is an eigenstate, are given in Tab. \ref{tab:matrix}.
       The analysis contemplates also the states of the doublet at 1000 {\icm}, since these
       states are coupled to some extent to the ones in the middle spectral-region, e.g.,
       the eigenstate $|\Psi_d^h\rangle$ has a squared overlap of $0.10$ with the test
       state $|\Phi_{bu}\rangle$.
       All states in Tab. \ref{tab:matrix} are of $B_2^+$ symmetry.

       State $|\Psi_{m1}\rangle$, absorbing at {\ESzARA} {\icm} in the MCTDH spectrum, 
       has a largest contribution from the $|\Phi_{1z,1R}\rangle$ test state. 
       The experimental {\zun}$\cdot$Ne spectrum shows a weak absorption in this region with
       a lower intensity than the peaks at 1763 and 1878 {\icm}, the same trend as in the MCTDH spectrum.
       Based on the general good agreement between the experimental and MCTDH spectra
       we propose that this weak absorption is mainly a combined excitation $(1z,1R)$.
       A band appearing at 1600 {\icm} in the MM/VCI spectrum \cite{ham05:244301} was assigned
       to the $(1z,1R)$ transition. However, the experimental spectrum of {\zun}$\cdot$Ne
       shows no absorption at 1600 {\icm} \cite{ham05:244301}.
       
       State $|\Psi_{m2}\rangle$, which is responsible for the absorption at {\ESbendOP} {\icm} in the
       MCTDH spectrum, has the largest contribution from the $|\Phi_{bu}\rangle$ test state.
       This peak is then mainly related to the {\em ungerade} water-bending, and has been
       already assigned in Ref. \onlinecite{ham05:244301} and a number of works. This peak
       can be assigned already from a standard normal-modes analysis, since its main
       contribution is from an internal motion of the water monomers,
       and less from the relative motions between them.
       However, it must not come as a surprise that
       the eigenstate $|\Psi_{m2}\rangle$ has a total
       squared overlap of $0.26$ with the test states containing one quanta of excitation in
       the proton transfer coordinate, namely $|\Phi_{1z}\rangle$, $|\Phi_{1z,1R}\rangle$
       and $|\Phi_{1z,2R}\rangle$. In a fashion similar to the coupling to the $w_3$ wagging
       motion, as the proton approaches one water molecule the equilibrium value of the 
       \mbox{H-O-H} angle shifts to a larger value because this water molecule acquires more {\hyd} character.
       Conversely, the water molecule at a larger distance of the central proton acquires
       {\wat} character and the angle \mbox{H-O-H} shifts to lower values.
       
       State $|\Psi_{m3}\rangle$ is responsible for the absorption at {\ESzARB} {\icm} in the
       MCTDH spectrum, the lowest intensity feature of the doublet centered at about 1800 {\icm}.
       This state has the largest overlap with the $|\Phi_{1z,2R}\rangle$ test-state. Comparison
       in position and intensity of this peak to the corresponding one in the 
       experimental {\zun}$\cdot$Ne spectrum suggests that the latest is mainly related
       to the ($1z,2R$) excitation.
       
       Analysis of the values in Tab. \ref{tab:matrix} shows that the eigenstates in the
       region 1400-1900 {\icm} are characterized by the asymmetric bending and combinations
       of the proton-transfer fundamental and water-water-stretch fundamental and first overtone,
       with important couplings between them. Such a coupling was already noted by Bowman and
       collaborators by analyzing the CI coefficients of their MM/VCI expansion \cite{ham05:244301}.
       However, the exact nature of each band could not be disentangled, and it was concluded
       that this ``is indicative of large couplings among various zero-order states in this region''.
       We show that, despite such couplings exist and play an important role in shaping the spectrum, 
       the different eigenstates involved retain each its particular character 
       (note that the largest numbers in Tab. \ref{tab:matrix} appear at the diagonal) 
       and can be assigned to well defined transitions.
       
    \subsection{2000-4000 {\icm} region}

       Symmetry analysis of the OH stretchings of the water molecules within the
       $\mathcal{G}_{16}$ group predicts four vibrational states with labels
       $A_1^+$, $B_2^+$ and $E^+$, and the corresponding torsional splitting 
       $A_1^-$, $B_2^-$ and $E^-$ states.
       The states-labeling simplifies to $A_1$, $B_2$ and $E$, respectively, 
       when adopting the $\mathcal{D}_{2d}$ point group, in which the internal rotation
       around $\alpha$ is treated as unfeasible. 
       The use of the smaller group $\mathcal{D}_{2d}$ is reasonable here because all the
       minus-states are dark in the linear IR-spectrum.
       The $A_1$ state
       corresponds to the symmetrical, {\em gerade} stretch, the $B_2$ state
       is the symmetrical, {\em ungerade} stretch and the $E$ states are the
       two asymmetrical stretches of {\em gerade} and {\em ungerade} type.
       We recall that the notation {\em gerade}/{\em ungerade} is used to indicate
       $+/-$ linear combinations of the motions of the two monomers, while
       symmetric/antisymmetric refers to the OH motions {\em within} each monomer.
       A symmetry analysis based on the $\mathcal{C}_2$ point group, the symmetry group
       of the absolute minimum, labels the symmetric stretches as $A$ and $B$ states and
       the asymmetric stretches again as $A$ and $B$ states. 
       The harmonic analysis results (see Tab \ref{tab:states}) yield
       the symmetric stretches separated by less than 10 {\icm} and appear at
       a lower energy than the asymmetric stretches. The asymmetric stretches, despite
       not being degenerate, are very close in energy, separated by less than 1 {\icm}. 
       Exact degeneracy of the asymmetric stretches is regained as a consequence of the 
       feasibility of the wagging-motions, leading to $\mathcal{D}_{2d}$ symmetry.

       The MCTDH spectrum in the water-stretching region reveals two absorptions.
       The lowest energy one is related exclusively to the $z$ component of the dipole, while the largest
       energy, most intense band arises exclusively from the perpendicular component (seen Fig.~\ref{fig:IR04000}).
       The lowest energy absorption is related to the symmetric, {\em ungerade} stretching,
       while the highest energy absorption is related to the degenerate asymmetric stretchings.
       The energy separation between symmetric and asymmetric stretchings is known to be of about 
       80 {\icm} as seen from different
       experiments and computations \cite{yeh89:7319,ham05:244301}, so these peaks can be resolved
       by a propagation of about 260 fs. The splitting between symmetric {\em gerade} and {\em ungerade} stretchings
       is expected to be less than 10 {\icm} according to harmonic results. Thus, a propagation
       of about 2100 fs would be needed to resolve these peaks, which is unfeasible in a reasonable amount
       of computer time for this system. In order to resolve
       the symmetric stretchings two initial test-states were prepared by diagonalization
       of mode-operators, the symmetric {\em gerade} ($|\Phi_{sg}\rangle$) and
       symmetric {\em ungerade} ($|\Phi_{su}\rangle$) stretchings.
       Both test states were propagated and their auto- and crosscorrelation functions analyzed
       by means of the FD method.
       FD analysis yields the symmetric, in-phase stretching at {\ESstrSIP} {\icm}, and the symmetric, out-of-phase
       counterpart at {\ESstrSOP} {\icm}. Propagation of a test state $|\Phi_{sa}\rangle$ 
       of the asymmetric, doubly degenerate
       stretchings yields the peak centered at {\ESstrA} {\icm}. These results are in excellent agreement to the
       bands observed in the experimental {\zun}$\cdot$Ne spectrum \cite{ham05:244301}.
       Note that the propagation of $|\Phi_{sa}\rangle$ yields a result in better agreement
       to experiment than the peak in Fig.~\ref{fig:IRparts} obtained by propagation of the 
       perpendicular component of the dipole.
       This is due to the fact that the wavefunction obtained from application of the dipole operator
       to the ground state is a coherent superposition of several eigenstates that are propagated together.
       The more eigenstates are to be coherently propagated, the more complex is the dynamics of the wavepacket
       and the less accurate becomes the propagation for each individual state. For this reason, the most accurate
       energies (e.g. values in Table \ref{tab:states}) are obtained from either {\em improved relaxation} to
       the desired eigenstate, if possible, or propagation of carefully prepared wavepackets (referred to as
       test states in this work) which have
       as much overlap as possible with the eigenstate of interest or a group of them.
       As a final remark, the spectrum arising from direct excitation of the $z$ coordinate (propagation of
       the $|\Phi_{1z}\rangle$ test state) yields all the peaks in the range 800-2000 {\icm} with almost the
       same relative intensities than compared to the propagation of $\mu_z|\Psi_0\rangle$. However, the propagation
       of $|\Phi_{1z}\rangle$ gives a completely flat spectrum in the region above 3000 {\icm}.
       This indicates that the excitation of the stretching motions must occur locally on each water molecule
       and does not depend on the excitation of the central proton. Moreover, the coupling of the 
       stretching motions to the central proton is very weak, as indicated by the splitting of only
       7 {\icm} between {\em gerade} and {\em ungerade} symmetric stretchings and indicates that the OH
       stretchings of {\zun} are robust to the displacement of the central proton, contrary to the
       situation seen for wagging and bending modes. Despite the excitation of the stretchings is local in each
       water monomer, the calculated band at {\ESstrSOP} {\icm}
       corresponds to the symmetric, {\em ungerade} mode of $B_2$ symmetry, since
       both water molecules are oriented in opposite directions with respect to the $z$-direction incident field.
       The symmetric, {\em gerade} mode is a dark state of $A_1$ symmetry.


\section{Summary and Conclusion}

  The infrared absorption spectrum of the Zundel cation (\zun) is calculated
  by the quantum-dynamical multiconfiguration time-dependent Hartree (MCTDH) method
  in the linear-absorption spectral-range 0 -- 4000 {\icm}.
  The energies of all fundamentals and several overtones related 
  to different motions of the system are reported. 
  The cation is considered in its full dimensionality (15D). A curvilinear set of coordinates
  is used to describe the configuration of the system. Details on the derivation of the
  kinetic energy operator used, which is exact for total angular momentum $J=0$, and on the
  representation of the potential energy surface are given in the companion paper \cite{ven07:paperI}.

  The lowest frequency part of the spectrum, 
  shows a strong absorption at about 100 {\icm}
  due to the combination of the fundamental wagging-modes $w_1$, which are $E^-$ degenerate states,
  and the internal-rotation state ($1\alpha$) of $A_1^-$ symmetry. The resulting pair of degenerate
  states is of $E^+$ symmetry and therefore bright.
  The two absorptions at 250 and 500 {\icm} are related to the combination state of the fundamental
  waggings and the ($3\alpha$) state, which results in a pair of $E^+$ degenerate states and to
  the rocking fundamentals, also of $E^+$ symmetry, respectively.
  These absorptions are related to the component of the field perpendicular to the water-water axis.
  The spectral region at about 1000 {\icm} presents a double-peak absorption 
  which is the most characteristic feature of the spectrum. 
  This double peak is seen to arise from the coupling of the proton-transfer
  motion with a combination state involving the $w_3$ wagging mode.
  The reduced probability density of the $w_3$ wagging mode projected onto the
  wagging coordinates is shown in Fig.~\ref{fig:dens2}b. This state presents four
  probability maxima, each of which corresponds to a water in pyramidal conformation (H$_2$O character)
  while the other is in planar conformation (H$_3^+$O character), and has an energy of {\ESwagC} {\icm}.
  This state alone absorbs light only very weakly (see Fig.~\ref{fig:IR04000}), but the state arising from the combination of $w_3$ and
  the water-water stretching ({\ESRA} {\icm}) reaches an energy close to the natural absorption
  of the proton-transfer at about 1000 {\icm}. 
  The coupling between these both states explains the
  doublet absorption at about 1000 {\icm}, which is interpreted as
  a Fermi resonance between the combination state ($1R$,$w_3$) and the proton-transfer fundamental ($1z$).

  The region between 1000 and 2000 {\icm} presents three main absorptions at
  {\ESzARA}, {\ESbendOP} and {\ESzARB} {\icm}. The peak with the highest intensity
  in this region is the peak at {\ESbendOP} {\icm} which corresponds 
  to the {\em ungerade} bending motion of the water moieties. The eigenstate at
  {\ESbendOP} {\icm} has in addition important contributions from the proton-transfer
  mode and proton transfer combined with the water-water stretching mode (see Tab. \ref{tab:matrix}).
  The eigenstates absorbing at {\ESzARA} and {\ESzARB} {\icm} are both
  described by one quantum on the proton-transfer mode plus one and two quanta excitations,
  respectively, on the water-water stretching mode. The nature of the two eigenstates of the
  doublet at 1000 {\icm} and the three eigenstates between 1000 and 2000 {\icm} is
  described by clearly defined motions (diagonal elements in Tab. \ref{tab:matrix}), 
  e.g. the asymmetric bending. However, they constitute a set of coupled states
  (non-diagonal elements in Tab. \ref{tab:matrix}) featuring the wagging, bending,
  water-water stretch and proton-transfer motions.

  The region above 3000 {\icm} presents the direct absorptions of the OH-stretching motions
  starting at about 3600 {\icm}. Symmetry analysis of the OH-stretching motions in the
  $\mathcal{D}_{2d}$ or $\mathcal{G}_{16}$ groups reveals that the symmetric {\em gerade}
  and {\em ungerade} stretching transform according to the $A_1$ and $B_2$ representations,
  respectively. The {\em ungerade}, $B_2$ state absorbs at {\ESstrSOP} {\icm} due to the
  $z$-component of the field. The {\em gerade}, $A_1$ state, which is dark, has an energy of
  {\ESstrSIP} {\icm}. The small energy splitting between {\em gerade} and {\em ungerade}
  states shows that the coupling of these motions to the central proton motion
  must be very weak. Furthermore, our analysis
  also points out that the excitation of the {\em ungerade}, $B_2$ state is completely independent
  of the central proton excitation, and is caused by interaction of the field with the local
  dipole of each monomer. Symmetry analysis reveals also that the asymmetric {\em gerade}
  and {\em ungerade} states are $E$ degenerate. They absorb at {\ESstrA} {\icm} due to the
  component of the field perpendicular to the water-water axis.

  The fact that the {\zun} cation may interconvert between several low energy barriers connecting
  equivalent minima had
  already been pointed out by Wales \cite{wal99:10403}, who showed that the correct symmetry group,
  because wagging and internal rotation motions are allowed, is the permutation-inversion group 
  $\mathcal{G}_{16}$ \cite{note:symm}.
  Our study shows that the symmetry analysis in $\mathcal{G}_{16}$ is necessary to understand some important
  features of the IR linear absorption spectrum. In addition, it has been shown in Paper I
  that already in the ground vibrational state there is non-neglible probability of
  crossing the internal-rotation barrier. However, the consideration of the cation in the
  more familiar $\mathcal{D}_{2d}$ point group may provide an adequate labelling of the vibrational motions of the system
  as long as the internal-rotation mode is not involved in the considered states.

  The reported calculations are in excellent agreement to the experimental measurements 
  of Ref. \cite{ham05:244301} on the predissociation spectrum of {\zun}$\cdot$Ne. 
  The discrepancy of the MCTDH energies reported in Tab. \ref{tab:states} with respect to
  the position of the measured bright bands lies always below 22 {\icm}, the average discrepance
  is 14 {\icm} (see Tab. \ref{tab:states}). Such a remarkable consistency between experiment and theory 
  along the whole spectral range represents, on the one hand, a validation of the underlying
  potential energy surface of Huang {\em et al.} \cite{hua05:044308} and of the mode-based
  cluster expansion of the potential \cite{ven07:paperI} used in the quantum-dynamical simulations,
  but is also a clear indication that a suitable set of coordinates was selected to tackle the problem \cite{ven07:paperI}.
  On the other hand, it provides a validation of the measurements on the {\zun}$\cdot$Ne cation
  by the predissociation technique, since this method gives access to the infrared linear-absorption regime
  without noticeable disturbances caused by the messenger atom.
  The only essential disagreement between experiment and theory is in the intensities. The intensities
  are not measured absolutely but, in comparison to the bright double-peak at 1000 {\icm}, the structures
  which appear between 1400 and 2000 {\icm} are too low by a factor of 3 when compared to the
  MCTDH simulation. Such a discrepancy does not occur in the spectrum
  of {\zun}$\cdot$Ar \cite{ham05:244301}, which displays a relative intensity in that region similar 
  to the MCTDH one despite an incorrect shape.
  Therefore we conclude that more investigations are necessary to determine the origin of
  the discrepancy in relative intensities between experiment and theory in some regions of the spectrum.

  The fact that the reported simulations are successful in obtaining accurate results for a system 
  of the dimensionality of the protonated water-dimer is to be attributed in great part to the
  MCTDH algorithm, in which not only the expansion coefficients, but also the
  orbitals (here SPFs), are variationally optimal. For a 15-dimensional system the use
  of only 4 basis functions per degree of freedom represents of the order of $10^9$  configurations.
  The largest calculations reported here consist of about $10^7$ configurations, while already
  good results are obtained by using as few as $10^5$ configurations (see Tab. 3 in Paper I).
  Such an {\em early} convergence of the MCTDH method becomes crucial as high-dimensional 
  problems are attempted. 
  
  The reported simulations constitute a new example of the ability of
  the MCTDH method to tackle high dimensional, complex molecular systems in a rigorous
  manner, and they open exciting perspectives for the simulation and understanding of even more
  complicated systems by means of accurate, non-trivial quantum-dynamical methods.
  Last but not least, they provide important information on the spectroscopy and dynamics of the hydrated proton.


\section{Acknowledgments}

  The authors thank Prof. J. Bowman for providing the potential-energy routine,
  M. Brill for the help with the parallelized code,
  and the Scientific Supercomputing Center Karlsruhe for generously providing computer time.
  O. V. is grateful to the Alexander von Humboldt Foundation for financial support.
  Travel support by the Deutsche Forschungsgemeinschaft (DFG) is also gratefully
  acknowledged.




\begin{thebibliography}{32}
\expandafter\ifx\csname natexlab\endcsname\relax\def\natexlab#1{#1}\fi
\expandafter\ifx\csname bibnamefont\endcsname\relax
  \def\bibnamefont#1{#1}\fi
\expandafter\ifx\csname bibfnamefont\endcsname\relax
  \def\bibfnamefont#1{#1}\fi
\expandafter\ifx\csname citenamefont\endcsname\relax
  \def\citenamefont#1{#1}\fi
\expandafter\ifx\csname url\endcsname\relax
  \def\url#1{\texttt{#1}}\fi
\expandafter\ifx\csname urlprefix\endcsname\relax\def\urlprefix{URL }\fi
\providecommand{\bibinfo}[2]{#2}
\providecommand{\eprint}[2][]{\url{#2}}

\bibitem[{\citenamefont{Marx et~al.}(1999)\citenamefont{Marx, Tuckerman,
  Hutter, and Parrinello}}]{mar99:601}
\bibinfo{author}{\bibfnamefont{D.}~\bibnamefont{Marx}},
  \bibinfo{author}{\bibfnamefont{M.}~\bibnamefont{Tuckerman}},
  \bibinfo{author}{\bibfnamefont{J.}~\bibnamefont{Hutter}}, \bibnamefont{and}
  \bibinfo{author}{\bibfnamefont{M.}~\bibnamefont{Parrinello}},
  \bibinfo{journal}{Nature} \textbf{\bibinfo{volume}{397}},
  \bibinfo{pages}{601} (\bibinfo{year}{1999}).

\bibitem[{\citenamefont{Agmon}(1999)}]{agm99:493}
\bibinfo{author}{\bibfnamefont{N.}~\bibnamefont{Agmon}}, \bibinfo{journal}{Isr.
  J. Chem.} \textbf{\bibinfo{volume}{39}}, \bibinfo{pages}{493 }
  (\bibinfo{year}{1999}).

\bibitem[{\citenamefont{Jiang et~al.}(2000)\citenamefont{Jiang, Wang, Chang,
  Lin, Lee, Niedner-{S}chatteburg, and Chang}}]{jia00:1398}
\bibinfo{author}{\bibfnamefont{J.-C.} \bibnamefont{Jiang}},
  \bibinfo{author}{\bibfnamefont{Y.-S.} \bibnamefont{Wang}},
  \bibinfo{author}{\bibfnamefont{H.-C.} \bibnamefont{Chang}},
  \bibinfo{author}{\bibfnamefont{S.~H.} \bibnamefont{Lin}},
  \bibinfo{author}{\bibfnamefont{Y.~T.} \bibnamefont{Lee}},
  \bibinfo{author}{\bibfnamefont{G.}~\bibnamefont{Niedner-{S}chatteburg}},
  \bibnamefont{and} \bibinfo{author}{\bibfnamefont{H.-C.} \bibnamefont{Chang}},
  \bibinfo{journal}{J. Am. Chem. Soc.} \textbf{\bibinfo{volume}{122}},
  \bibinfo{pages}{1398 } (\bibinfo{year}{2000}).

\bibitem[{\citenamefont{Headrick et~al.}(2005)\citenamefont{Headrick, Diken,
  Walters, Hammer, Christie, Cui, Myshakin, Duncan, Johnson, and
  Jordan}}]{hea05:1765}
\bibinfo{author}{\bibfnamefont{J.~M.} \bibnamefont{Headrick}},
  \bibinfo{author}{\bibfnamefont{E.~G.} \bibnamefont{Diken}},
  \bibinfo{author}{\bibfnamefont{R.~S.} \bibnamefont{Walters}},
  \bibinfo{author}{\bibfnamefont{N.~I.} \bibnamefont{Hammer}},
  \bibinfo{author}{\bibfnamefont{R.~A.} \bibnamefont{Christie}},
  \bibinfo{author}{\bibfnamefont{J.}~\bibnamefont{Cui}},
  \bibinfo{author}{\bibfnamefont{E.~M.} \bibnamefont{Myshakin}},
  \bibinfo{author}{\bibfnamefont{M.~A.} \bibnamefont{Duncan}},
  \bibinfo{author}{\bibfnamefont{M.~A.} \bibnamefont{Johnson}},
  \bibnamefont{and} \bibinfo{author}{\bibfnamefont{K.~D.}
  \bibnamefont{Jordan}}, \bibinfo{journal}{Science}
  \textbf{\bibinfo{volume}{308}}, \bibinfo{pages}{1765} (\bibinfo{year}{2005}).

\bibitem[{\citenamefont{Hammer et~al.}(2005)\citenamefont{Hammer, Diken,
  Roscioli, Johnson, Myshakin, Jordan, Mc{C}oy, Huang, Bowman, and
  Carter}}]{ham05:244301}
\bibinfo{author}{\bibfnamefont{N.~I.} \bibnamefont{Hammer}},
  \bibinfo{author}{\bibfnamefont{E.~G.} \bibnamefont{Diken}},
  \bibinfo{author}{\bibfnamefont{J.~R.} \bibnamefont{Roscioli}},
  \bibinfo{author}{\bibfnamefont{M.~A.} \bibnamefont{Johnson}},
  \bibinfo{author}{\bibfnamefont{E.~M.} \bibnamefont{Myshakin}},
  \bibinfo{author}{\bibfnamefont{K.~D.} \bibnamefont{Jordan}},
  \bibinfo{author}{\bibfnamefont{A.~B.} \bibnamefont{Mc{C}oy}},
  \bibinfo{author}{\bibfnamefont{X.}~\bibnamefont{Huang}},
  \bibinfo{author}{\bibfnamefont{J.~M.} \bibnamefont{Bowman}},
  \bibnamefont{and} \bibinfo{author}{\bibfnamefont{S.}~\bibnamefont{Carter}},
  \bibinfo{journal}{J. Chem. Phys.} \textbf{\bibinfo{volume}{122}},
  \bibinfo{pages}{244301} (\bibinfo{year}{2005}).

\bibitem[{\citenamefont{Asmis et~al.}(2003)\citenamefont{Asmis, Pivonka,
  Santambrogio, Brummer, Kaposta, Neumark, and Woste}}]{asm03:1375}
\bibinfo{author}{\bibfnamefont{K.~R.} \bibnamefont{Asmis}},
  \bibinfo{author}{\bibfnamefont{N.~L.} \bibnamefont{Pivonka}},
  \bibinfo{author}{\bibfnamefont{G.}~\bibnamefont{Santambrogio}},
  \bibinfo{author}{\bibfnamefont{M.}~\bibnamefont{Brummer}},
  \bibinfo{author}{\bibfnamefont{C.}~\bibnamefont{Kaposta}},
  \bibinfo{author}{\bibfnamefont{D.~M.} \bibnamefont{Neumark}},
  \bibnamefont{and} \bibinfo{author}{\bibfnamefont{L.}~\bibnamefont{Woste}},
  \bibinfo{journal}{Science} \textbf{\bibinfo{volume}{299}},
  \bibinfo{pages}{1375 } (\bibinfo{year}{2003}).

\bibitem[{\citenamefont{Fridgen et~al.}(2004)\citenamefont{Fridgen, Mc{M}ahon,
  Mac{A}leese, Lemaire, and Maitre}}]{fri04:9008}
\bibinfo{author}{\bibfnamefont{T.~D.} \bibnamefont{Fridgen}},
  \bibinfo{author}{\bibfnamefont{T.~B.} \bibnamefont{Mc{M}ahon}},
  \bibinfo{author}{\bibfnamefont{L.}~\bibnamefont{Mac{A}leese}},
  \bibinfo{author}{\bibfnamefont{J.}~\bibnamefont{Lemaire}}, \bibnamefont{and}
  \bibinfo{author}{\bibfnamefont{P.}~\bibnamefont{Maitre}},
  \bibinfo{journal}{J. Phys. Chem. A} \textbf{\bibinfo{volume}{108}},
  \bibinfo{pages}{9008 } (\bibinfo{year}{2004}).

\bibitem[{\citenamefont{Headrick et~al.}(2004)\citenamefont{Headrick, Bopp, and
  Johnson}}]{hea04:11523}
\bibinfo{author}{\bibfnamefont{J.~M.} \bibnamefont{Headrick}},
  \bibinfo{author}{\bibfnamefont{J.~C.} \bibnamefont{Bopp}}, \bibnamefont{and}
  \bibinfo{author}{\bibfnamefont{M.~A.} \bibnamefont{Johnson}},
  \bibinfo{journal}{J. Chem. Phys} \textbf{\bibinfo{volume}{121}},
  \bibinfo{pages}{11523} (\bibinfo{year}{2004}).

\bibitem[{\citenamefont{Vener et~al.}(2001)\citenamefont{Vener, K{\"u}hn, and
  Sauer}}]{ven01:240}
\bibinfo{author}{\bibfnamefont{M.~V.} \bibnamefont{Vener}},
  \bibinfo{author}{\bibfnamefont{O.}~\bibnamefont{K{\"u}hn}}, \bibnamefont{and}
  \bibinfo{author}{\bibfnamefont{J.}~\bibnamefont{Sauer}}, \bibinfo{journal}{J.
  Chem. Phys.} \textbf{\bibinfo{volume}{114}}, \bibinfo{pages}{240 }
  (\bibinfo{year}{2001}).

\bibitem[{\citenamefont{Dai et~al.}(2003)\citenamefont{Dai, Bacic, Huang,
  Carter, and Bowman}}]{dai03:6571}
\bibinfo{author}{\bibfnamefont{J.}~\bibnamefont{Dai}},
  \bibinfo{author}{\bibfnamefont{Z.}~\bibnamefont{Bacic}},
  \bibinfo{author}{\bibfnamefont{X.~C.} \bibnamefont{Huang}},
  \bibinfo{author}{\bibfnamefont{S.}~\bibnamefont{Carter}}, \bibnamefont{and}
  \bibinfo{author}{\bibfnamefont{J.~M.} \bibnamefont{Bowman}},
  \bibinfo{journal}{J. Chem. Phys.} \textbf{\bibinfo{volume}{119}},
  \bibinfo{pages}{6571 } (\bibinfo{year}{2003}).

\bibitem[{\citenamefont{Huang et~al.}(2005)\citenamefont{Huang, Braams, and
  Bowman}}]{hua05:044308}
\bibinfo{author}{\bibfnamefont{X.}~\bibnamefont{Huang}},
  \bibinfo{author}{\bibfnamefont{B.~J.} \bibnamefont{Braams}},
  \bibnamefont{and} \bibinfo{author}{\bibfnamefont{J.~M.}
  \bibnamefont{Bowman}}, \bibinfo{journal}{J. Chem. Phys.}
  \textbf{\bibinfo{volume}{122}}, \bibinfo{pages}{044308}
  (\bibinfo{year}{2005}).

\bibitem[{\citenamefont{Sauer and Dobler}(2005)}]{sau05:1706}
\bibinfo{author}{\bibfnamefont{J.}~\bibnamefont{Sauer}} \bibnamefont{and}
  \bibinfo{author}{\bibfnamefont{J.}~\bibnamefont{Dobler}},
  \bibinfo{journal}{Chem. Phys. Chem.} \textbf{\bibinfo{volume}{6}},
  \bibinfo{pages}{1706 } (\bibinfo{year}{2005}).

\bibitem[{\citenamefont{Kaledin et~al.}(2006)\citenamefont{Kaledin, Kaledin,
  and Bowman}}]{kal06:2933}
\bibinfo{author}{\bibfnamefont{M.}~\bibnamefont{Kaledin}},
  \bibinfo{author}{\bibfnamefont{A.~L.} \bibnamefont{Kaledin}},
  \bibnamefont{and} \bibinfo{author}{\bibfnamefont{J.~M.}
  \bibnamefont{Bowman}}, \bibinfo{journal}{J. Phys. Chem. A}
  \textbf{\bibinfo{volume}{110}}, \bibinfo{pages}{2933 }
  (\bibinfo{year}{2006}).

\bibitem[{\citenamefont{Vendrell
  et~al.}(2007{\natexlab{a}})\citenamefont{Vendrell, Gatti, and
  Meyer}}]{ven07:letter}
\bibinfo{author}{\bibfnamefont{O.}~\bibnamefont{Vendrell}},
  \bibinfo{author}{\bibfnamefont{F.}~\bibnamefont{Gatti}}, \bibnamefont{and}
  \bibinfo{author}{\bibfnamefont{H.-D.} \bibnamefont{Meyer}},
  \bibinfo{journal}{Angewandte Chemie} \textbf{\bibinfo{volume}{in press}}
  (\bibinfo{year}{2007}{\natexlab{a}}).

\bibitem[{\citenamefont{Vendrell
  et~al.}(2007{\natexlab{b}})\citenamefont{Vendrell, Gatti, Lauvergnat, and
  Meyer}}]{ven07:paperI}
\bibinfo{author}{\bibfnamefont{O.}~\bibnamefont{Vendrell}},
  \bibinfo{author}{\bibfnamefont{F.}~\bibnamefont{Gatti}},
  \bibinfo{author}{\bibfnamefont{D.}~\bibnamefont{Lauvergnat}},
  \bibnamefont{and} \bibinfo{author}{\bibfnamefont{H.-D.} \bibnamefont{Meyer}},
  \bibinfo{journal}{J. Chem. Phys.}  (\bibinfo{year}{2007}{\natexlab{b}}).

\bibitem[{\citenamefont{Manthe et~al.}(1992)\citenamefont{Manthe, Meyer, and
  Cederbaum}}]{man92:3199}
\bibinfo{author}{\bibfnamefont{U.}~\bibnamefont{Manthe}},
  \bibinfo{author}{\bibfnamefont{H.-D.} \bibnamefont{Meyer}}, \bibnamefont{and}
  \bibinfo{author}{\bibfnamefont{L.~S.} \bibnamefont{Cederbaum}},
  \bibinfo{journal}{J.~Chem.\ Phys.} \textbf{\bibinfo{volume}{97}},
  \bibinfo{pages}{3199} (\bibinfo{year}{1992}).

\bibitem[{\citenamefont{Beck et~al.}(2000)\citenamefont{Beck, J{\"a}ckle,
  Worth, and Meyer}}]{bec00:1}
\bibinfo{author}{\bibfnamefont{M.~H.} \bibnamefont{Beck}},
  \bibinfo{author}{\bibfnamefont{A.}~\bibnamefont{J{\"a}ckle}},
  \bibinfo{author}{\bibfnamefont{G.~A.} \bibnamefont{Worth}}, \bibnamefont{and}
  \bibinfo{author}{\bibfnamefont{H.-D.} \bibnamefont{Meyer}},
  \bibinfo{journal}{Phys.\ Rep.} \textbf{\bibinfo{volume}{324}},
  \bibinfo{pages}{1} (\bibinfo{year}{2000}).

\bibitem[{\citenamefont{Meyer and Worth}(2003)}]{mey03:251}
\bibinfo{author}{\bibfnamefont{H.-D.} \bibnamefont{Meyer}} \bibnamefont{and}
  \bibinfo{author}{\bibfnamefont{G.~A.} \bibnamefont{Worth}},
  \bibinfo{journal}{Theor.\ Chem.\ Acc.} \textbf{\bibinfo{volume}{109}},
  \bibinfo{pages}{251} (\bibinfo{year}{2003}).

\bibitem[{\citenamefont{Worth et~al.}()\citenamefont{Worth, Beck, J{\"a}ckle,
  and Meyer}}]{mctdh:package}
\bibinfo{author}{\bibfnamefont{G.~A.} \bibnamefont{Worth}},
  \bibinfo{author}{\bibfnamefont{M.~H.} \bibnamefont{Beck}},
  \bibinfo{author}{\bibfnamefont{A.}~\bibnamefont{J{\"a}ckle}},
  \bibnamefont{and} \bibinfo{author}{\bibfnamefont{H.-D.} \bibnamefont{Meyer}},
  \bibinfo{howpublished}{The {MCTDH} {P}ackage, {V}ersion 8.2, (2000). H.-D.
  Meyer, {V}ersion 8.3 (2002), {V}ersion 8.4 (2007). {S}ee
  http://www.pci.uni-heidelberg.de/tc/usr/mctdh/}.

\bibitem[{\citenamefont{Balint-Kurti et~al.}(1990)\citenamefont{Balint-Kurti,
  Dixon, and Marston}}]{bal90:1741}
\bibinfo{author}{\bibfnamefont{G.~G.} \bibnamefont{Balint-Kurti}},
  \bibinfo{author}{\bibfnamefont{R.~N.} \bibnamefont{Dixon}}, \bibnamefont{and}
  \bibinfo{author}{\bibfnamefont{C.~C.} \bibnamefont{Marston}},
  \bibinfo{journal}{J.~Chem.\ Soc., Faraday Trans.}
  \textbf{\bibinfo{volume}{86}}, \bibinfo{pages}{1741} (\bibinfo{year}{1990}).

\bibitem[{\citenamefont{Meyer et~al.}(2006)\citenamefont{Meyer, {Le Qu\'er\'e},
  L\'eonard, and Gatti}}]{mey06:179}
\bibinfo{author}{\bibfnamefont{H.-D.} \bibnamefont{Meyer}},
  \bibinfo{author}{\bibfnamefont{F.}~\bibnamefont{{Le Qu\'er\'e}}},
  \bibinfo{author}{\bibfnamefont{C.}~\bibnamefont{L\'eonard}},
  \bibnamefont{and} \bibinfo{author}{\bibfnamefont{F.}~\bibnamefont{Gatti}},
  \bibinfo{journal}{Chem.\ Phys.} \textbf{\bibinfo{volume}{329}},
  \bibinfo{pages}{179} (\bibinfo{year}{2006}).

\bibitem[{\citenamefont{Richter et~al.}(2007)\citenamefont{Richter, Gatti,
  L{\'e}onard, {Le {Q}u{\'e}r{\'e}}, and Meyer}}]{ric07:xxx}
\bibinfo{author}{\bibfnamefont{F.}~\bibnamefont{Richter}},
  \bibinfo{author}{\bibfnamefont{F.}~\bibnamefont{Gatti}},
  \bibinfo{author}{\bibfnamefont{C.}~\bibnamefont{L{\'e}onard}},
  \bibinfo{author}{\bibfnamefont{F.}~\bibnamefont{{Le {Q}u{\'e}r{\'e}}}},
  \bibnamefont{and} \bibinfo{author}{\bibfnamefont{H.-D.} \bibnamefont{Meyer}},
  \bibinfo{journal}{J. Chem. Phys.} \textbf{\bibinfo{volume}{in press}}
  (\bibinfo{year}{2007}).

\bibitem[{\citenamefont{Wall and Neuhauser}(1995)}]{wal95:8011}
\bibinfo{author}{\bibfnamefont{M.~R.} \bibnamefont{Wall}} \bibnamefont{and}
  \bibinfo{author}{\bibfnamefont{D.}~\bibnamefont{Neuhauser}},
  \bibinfo{journal}{J.~Chem.\ Phys.} \textbf{\bibinfo{volume}{102}},
  \bibinfo{pages}{8011} (\bibinfo{year}{1995}).

\bibitem[{\citenamefont{Beck and Meyer}(1998)}]{bec98:3730}
\bibinfo{author}{\bibfnamefont{M.~H.} \bibnamefont{Beck}} \bibnamefont{and}
  \bibinfo{author}{\bibfnamefont{H.-D.} \bibnamefont{Meyer}},
  \bibinfo{journal}{J.~Chem.\ Phys.} \textbf{\bibinfo{volume}{109}},
  \bibinfo{pages}{3730} (\bibinfo{year}{1998}).

\bibitem[{\citenamefont{Beck and Meyer}(2001)}]{bec01:2036}
\bibinfo{author}{\bibfnamefont{M.~H.} \bibnamefont{Beck}} \bibnamefont{and}
  \bibinfo{author}{\bibfnamefont{H.-D.} \bibnamefont{Meyer}},
  \bibinfo{journal}{J.~Chem.\ Phys.} \textbf{\bibinfo{volume}{114}},
  \bibinfo{pages}{2036} (\bibinfo{year}{2001}).

\bibitem[{\citenamefont{Gatti et~al.}(2001)\citenamefont{Gatti, Beck, Worth,
  and Meyer}}]{gat01:1576}
\bibinfo{author}{\bibfnamefont{F.}~\bibnamefont{Gatti}},
  \bibinfo{author}{\bibfnamefont{M.~H.} \bibnamefont{Beck}},
  \bibinfo{author}{\bibfnamefont{G.~A.} \bibnamefont{Worth}}, \bibnamefont{and}
  \bibinfo{author}{\bibfnamefont{H.-D.} \bibnamefont{Meyer}},
  \bibinfo{journal}{PCCP} \textbf{\bibinfo{volume}{3}}, \bibinfo{pages}{1576}
  (\bibinfo{year}{2001}).

\bibitem[{\citenamefont{{Longuet-{H}iggins}}(1963)}]{lon63:445}
\bibinfo{author}{\bibfnamefont{H.~C.} \bibnamefont{{Longuet-{H}iggins}}},
  \bibinfo{journal}{Mol. Phys.} \textbf{\bibinfo{volume}{6}},
  \bibinfo{pages}{445 } (\bibinfo{year}{1963}).

\bibitem[{\citenamefont{Bunker and Jensen}(1998)}]{bunker-book:symmetry}
\bibinfo{author}{\bibfnamefont{P.~R.} \bibnamefont{Bunker}} \bibnamefont{and}
  \bibinfo{author}{\bibfnamefont{P.}~\bibnamefont{Jensen}},
  \emph{\bibinfo{title}{Molecular Symmetry and Spectroscopy}}
  (\bibinfo{publisher}{NRC Research Press}, \bibinfo{year}{1998}),
  \bibinfo{edition}{2nd} ed.

\bibitem[{\citenamefont{Wales}(1999)}]{wal99:10403}
\bibinfo{author}{\bibfnamefont{D.~J.} \bibnamefont{Wales}},
  \bibinfo{journal}{J. Chem. Phys.} \textbf{\bibinfo{volume}{110}},
  \bibinfo{pages}{10403 } (\bibinfo{year}{1999}).

\bibitem[{\citenamefont{Bowman et~al.}(2003)\citenamefont{Bowman, Carter, and
  Huang}}]{bow03:533}
\bibinfo{author}{\bibfnamefont{J.~M.} \bibnamefont{Bowman}},
  \bibinfo{author}{\bibfnamefont{S.}~\bibnamefont{Carter}}, \bibnamefont{and}
  \bibinfo{author}{\bibfnamefont{X.}~\bibnamefont{Huang}},
  \bibinfo{journal}{Int. Rev. Phys. Chem.} \textbf{\bibinfo{volume}{22}},
  \bibinfo{pages}{533} (\bibinfo{year}{2003}).

\bibitem[{\citenamefont{Yeh et~al.}(1989)\citenamefont{Yeh, Okumura, Myers,
  Price, and Lee}}]{yeh89:7319}
\bibinfo{author}{\bibfnamefont{L.~I.} \bibnamefont{Yeh}},
  \bibinfo{author}{\bibfnamefont{M.}~\bibnamefont{Okumura}},
  \bibinfo{author}{\bibfnamefont{J.~D.} \bibnamefont{Myers}},
  \bibinfo{author}{\bibfnamefont{J.~M.} \bibnamefont{Price}}, \bibnamefont{and}
  \bibinfo{author}{\bibfnamefont{Y.~T.} \bibnamefont{Lee}},
  \bibinfo{journal}{The Journal of Chemical Physics}
  \textbf{\bibinfo{volume}{91}}, \bibinfo{pages}{7319} (\bibinfo{year}{1989}),

\bibitem[{\citenamefont{Mc{C}oy et~al.}(2005)\citenamefont{Mc{C}oy, Huang,
  Carter, Landeweer, and Bowman}}]{mc05:061101}
\bibinfo{author}{\bibfnamefont{A.~B.} \bibnamefont{Mc{C}oy}},
  \bibinfo{author}{\bibfnamefont{X.}~\bibnamefont{Huang}},
  \bibinfo{author}{\bibfnamefont{S.}~\bibnamefont{Carter}},
  \bibinfo{author}{\bibfnamefont{M.~Y.} \bibnamefont{Landeweer}},
  \bibnamefont{and} \bibinfo{author}{\bibfnamefont{J.~M.}
  \bibnamefont{Bowman}}, \bibinfo{journal}{J. Chem. Phys.}
  \textbf{\bibinfo{volume}{122}}, \bibinfo{pages}{061101}
  (\bibinfo{year}{2005}).

\bibitem[{\citenamefont{N}(2007)\citenamefont{N}}]{note:symm}
\bibinfo{author}{Note that Wales \cite{wal99:10403} uses a convention to
assign irreducible representations which differs from the one we use \cite{bunker-book:symmetry}.
The latter convention has the advantage to be consistent with the convention 
of the $\mathcal{D}_{2d}$ point group.}

\end{thebibliography}

\clearpage
  \begin{table}
    {\tiny
    \caption{\label{tab:states} Vibrational-excited states as identified in the MCTDH calculations.
                                Comparison is given to harmonic-analysis (HO) results in the same surface,
                                the MM/VCI results and the experimental results on the {\zun}$\cdot$Ne
                                cation. In the MCTDH column the subscript $D$ indicates that the state
                                was obtained by {\em improved relaxation} and the subscript $F$ indicates
                                that the state was identified by Fourier analysis.
                                In the $|\langle\Psi_n|\hat{\mu}|\Psi_0\rangle|^2$ column, $\hat{\mu}$ refers
                                to $\hat{\mu}_z$ for $B_2^+$ states and to $\hat{\mu}_x$ (or $\hat{\mu}_y$)
                                for $E^+$ states.
                                In the assignments column, in parenthesis, a number followed by a letter 
                                indicates the quanta of excitation in that coordinate. 
                                Other states are named after their definition in the text.
                                As a remainder we note that $w$, $r$, $b$ and $s$ refer to
                                wagging, rocking, bending and stretching.
                                When meaningful a {\em ket} description of the state is given.
                                In the {\em ket} description of the OH stretchings, $S/A$ 
                                indicate symmetric/asymmetric stretching-motion within each monomer, respectively.
                                The last column shows the irreducible representation of the $\mathcal{G}_{16}$
                                permutation-inversion group to which the vibrational state belongs.
                                The irreducible representations of the more familiar 
                                point-group $\mathcal{D}_{2d}$, which is a subgroup of $\mathcal{G}_{16}$, are
                                obtained by simply dropping the upper ($+/-$) index.
                                }
    \begin{tabular}{c|cccccc|cc}
        \hline
        \hline
                      Description       
                  &   HO~\footnotemark[1]    
                  &   VCI(DMC)~\footnotemark[2]
                  &   Exp.~\footnotemark[3]
                  &   MCTDH~\footnotemark[4]
                  &   \multicolumn{2}{c|}{assignment}
                  &   $|\langle\Psi_n|\hat{\mu}|\Psi_0\rangle|^2$
                  &   $\mathcal{G}_{16}$                              \\
        \hline

                  &   
                  &                  
                  &
                  &   {\ESrotO}$_{(D)}$    &  $(1\alpha)$  & 
                  &   dark                    
                  &   $A_1^-$                         \\
                      Torsion
                  &   170      
                  &                  
                  &
                  &   {\ESrotA}$_{(D)}$    &  $(2\alpha)$  &
                  &   dark
                  &   $B_1^+$                         \\

                  & 
                  &                  
                  &
                  &   {\ESrotB}$_{(D)}$    &  $(3\alpha)$  &
                  &   dark
                  &   $B_1^-$                         \\

                  & 
                  &                  
                  &
                  &   {\ESrotC}$_{(F)}$    &  $(4\alpha)$  &
                  &   dark
                  &   $A_1^+$                         \\
        \hline
                  &   339, 471 
                  &   
                  &
                  &   {\ESwagA}$_{(D)}$    &  $w_{1a,b}$ &  $|10\rangle \pm |01\rangle$
                  &   dark
                  &   $E^-$                           \\

                  &   
                  &   
                  &
                  &   {\ESwagAaa}$_{(D)}$    &  $(1\alpha,w_{1a,b})$ & 
                  &   0.126
                  &   $E^+$                           \\
                      Wagging
                  &   
                  &   
                  &
                  &   {\ESwagB}$_{(D)}$    &  $w_{2}$ & $|11\rangle$ 
                  &   dark
                  &   $B_1^+$                                \\

                  &   
                  &   
                  &
                  &   {\ESwagAab}$_{(F)}$    &  $(3\alpha,w_{1a,b})$ & 
                  &   0.010
                  &   $E^+$                           \\
                      
                  &   
                  &   
                  &
                  &   {\ESwagC}$_{(D)}$    & $w_{3}$ & $|20\rangle - |02\rangle$
                  &   0.0017
                  &   $B_2^+$                         \\
                     
                  &   
                  &   
                  &
                  &   {\ESwagD}$_{(D)}$    & $w_{4}$ & $|20\rangle + |02\rangle$
                  &   dark
                  &   $A_1^+$                         \\
        \hline
                  &   532, 554 
                  &                  
                  &
                  &   {\ESrock}$_{(F)}$    &  $r_{1a,b}$ &  $|10\rangle \pm |01\rangle$ 
                  &   0.0021
                  &   $E^+$                           \\
                      Rocking
                  &   
                  &                  
                  &
                  &   {\ESrockD}$_{(F)}$    & $r_{2}$ &  $|20\rangle + |02\rangle$
                  &   dark
                  &   $A_1^+$                           \\
                      
                  &   
                  &                  
                  &
                  &   {\ESrockC}$_{(F)}$    & $r_{3}$ & $|20\rangle - |02\rangle$
                  &   0.0071
                  &   $B_2^+$                            \\
                      
                  &   
                  &                  
                  &
                  &   {\ESrockB}$_{(F)}$    & $r_{4}$ & $|11\rangle$
                  &   dark
                  &   $B_1^+$                             \\
        \hline
                      Wat-Wat str.
                  &   630      
                  &                  
                  &
                  &   {\ESRA}$_{(D)}$      &   ($1R$) &
                  &   dark
                  &   $A_1^+$                         \\
                                  
                  &         
                  &                  
                  &
                  &   {\ESRB}$_{(D)}$      &  ($2R$) &
                  &   dark
                  &   $A_1^+$                          \\
        \hline
                      Proton-transfer  
                  &         
                  &   
                  &   928
                  &   {\ESdoubl}$_{(F)}$ &  ($1R,w_3$) &
                  &   0.042
                  &   $B_2^+$                         \\
                      doublet
                  &   861
                  &   1070(995)
                  &   1047
                  &   {\ESdoubh}$_{(F)}$  &   ($1z$) &
                  &   0.079 
                  &   $B_2^+$                         \\
                     and overtone 
                  &   
                  &   
                  &   
                  &   {\ESzB}$_{(F)}$   & ($2z$)  &  
                  &   dark
                  &   $A_1^+$                         \\
        \hline

        \hline
                      Proton perp.
                  &   1494, 1574         
                  &   
                  &
                  &   {\ESperpP}$_{(F)}$  &   ($1x$),($1y$) &
                  &   0.00076
                  &   $E^{+}$                         \\
        \hline
                      Proton-transfer +
                  &   
                  &   1600
                  &   $\approx$ 1425  
                  &   {\ESzARA}$_{(F)}$  &   ($1z,1R$) &
                  &   0.0064
                  &   $B_2^+$                         \\
                      Wat-Wat str.
                  &   
                  &   1832/1910
                  &   1878
                  &   {\ESzARB}$_{(F)}$  &  ($1z,2R$) &
                  &   0.0063
                  &   $B_2^+$                         \\
        \hline
                      Wat. bend ({\em gerade})              
                  &   1720     
                  &   1604               
                  &
                  &   {\ESbendIP}$_{(F)}$  & $bg$   &  $|10\rangle + |01\rangle$
                  &   dark
                  &   $A_1^+$                         \\
                      \phantom{Wat. bend }({\em ungerade})
                  &   1770     
                  &   1781               
                  &   1763
                  &   {\ESbendOP}$_{(F)}$   & $bu$  &  $|10\rangle - |01\rangle$
                  &   0.019
                  &   $B_2^+$                         \\
        \hline
                      O-H (sym,{\em gerade})   
                  &   3744     
                  &   3610(3511)      
                  &   
                  &   {\ESstrSIP}$_{(F)}$    & $sg$   & $|S0\rangle + |0S\rangle$
                  &   dark
                  &   $A_1^+$                         \\
            \phantom{O-H }(sym,{\em ungerade}) 
                  &   3750     
                  &   3625(3553) 
                  &   3603
                  &   {\ESstrSOP}$_{(F)}$    & $su$   & $|S0\rangle - |0S\rangle$
                  &   0.0028
                  &   $B_2^+$                         \\
             \phantom{O-H }(asym)                                  
                  &   3832   
                  &   3698(3652) 
                  &   3683  
                  &   {\ESstrA}$_{(F)}$     & $sa$   & $|A0\rangle \pm |0A\rangle$
                  &   0.0028
                  &   $E^+$                           \\
        \hline
        \hline
    \end{tabular}
        \footnotetext[1]{Normal mode harmonic analysis. Results are taken from Ref. \cite{hua05:044308}.}
        \footnotetext[2]{Energies computed by the MULTIMODE program and (in parenthesis) by
                         diffusion Monte-Carlo. Results taken grom Ref. \cite{mc05:061101}.}
        \footnotetext[3]{Experimental results taken from Ref. \cite{mc05:061101}.}
        \footnotetext[4]{This work.}
    }
  \end{table}

\clearpage
       \begin{table}
          \caption{\label{tab:matrix} Overlaps $|\langle\Phi|\Psi\rangle|^2$, 
                          where $|\Phi\rangle$ are test states and $|\Psi\rangle$ are eigenstates.
                          }
           \begin{tabular}{cc@{\hspace{0.5cm}}c@{\hspace{0.5cm}}c@{\hspace{0.5cm}}c@{\hspace{0.5cm}}c}
            \hline
            \hline
              & $|\Psi_d^l\rangle$ & $|\Psi_d^h\rangle$ & $|\Psi_{m1}\rangle$ & $|\Psi_{m2}\rangle$ & $|\Psi_{m3}\rangle$ \\
            \hline
Energy [\icm] &  \ESdoubl          & \ESdoubh           & \ESzARA             & \ESbendOP           & \ESzARB  \\
            \hline
$\langle\Phi_{1R,w_3}|$ & 0.83       &  0.09            & 0.00                & 0.00                & 0.00            \\
$\langle\Phi_{1z}|$     & 0.10       &  0.46            & 0.04                & 0.10                & 0.04            \\
$\langle\Phi_{1z,1R}|$  & 0.00       &  0.00            & 0.44                & 0.06                & 0.01            \\
$\langle\Phi_{bu}|$     & 0.07       &  0.12            & 0.10                & 0.38                & 0.02            \\
$\langle\Phi_{1z,2R}|$  & 0.00       &  0.00            & 0.00                & 0.10                & 0.38            \\
            \hline
            \hline
           \end{tabular}
       \end{table}

\clearpage
\section*{Figure Captions}

    \figcaption{fig:IR04000}{Simulated MCTDH spectrum in the range 0--4000 {\icm}. 
              Excitation in the $z$ direction (top), perpendicular
              to the O-H-O axis (middle) and total spectrum, i.e. $(1/3)z$ + $(2/3)*$perpendicular (bottom). 
              Note the different scale of intensities in the perpendicular-component plot.
              Autocorrelation time $T=1000$ fs.
              Absorption is given in absolute
              scale in mega-barns (Mb). (1 Mb = 10$^{-18}$ cm$^2$).}

    \figcaption{fig:IRparts}{Comparison between the MCTDH spectrum (top) and the \zun$\cdot$Ne spectrum of 
              Ref. \cite{ham05:244301} (bottom). 
              The intensity of the experimental spectrum is adjusted in each spectral region
              (800-2000 and 3500-3800 {\icm}) using the most intense peak of the MCTDH spectrum as a reference.
              Absorption for the MCTDH spectrum is given in absolute
              scale in mega-barns (Mb). (1 Mb = 10$^{-18}$ cm$^2$).}

    \figcaption{fig:dens3}{Reduced probability density on the internal rotation $\alpha$
              for (a) the ground state and the first three excited states:
              (b) $1\alpha$, (c) $2\alpha$ and (d) $3\alpha$. 
              The dotted lines correspond to an enlarged scale ($\times 10$).
              The $+/-$ symbols are intended to clarify the symmetry properties of
              each state. They indicate the sign of the underlying wavefunction
              based on a 1D computation for coordinate $\alpha$ and do not refer directly
              to the multidimensional wavefunctions from which densities are given.}

    \figcaption{fig:dens1}{Reduced probability density on the wagging coordinates $\gamma_A$ and $\gamma_B$
             of (a) the ground vibrational state and
             (b) the first-excited ($w_{1a}$) wagging-mode states.}

    \figcaption{fig:dens2}{Reduced probability density on the wagging coordinates $\gamma_A$ and $\gamma_B$
              of the excited states $w_2$, $w_3$ and $w_4$, characterized by two quanta of excitation.
              Compare with Table \ref{tab:states}.}

    \figcaption{fig:pictorial}{Schematic representation of the two most important 
                  coupled motions responsible for the doublet peak at 1000 {\icm}
              }

    \clearpage

    \clearpage

    \clearpage

    \clearpage

    \clearpage

    \clearpage


\end{document}